\documentstyle[twocolumn,prd,aps]{revtex}
\input{epsf}
\begin{document}
%
%
\title{RENORMALIZABILITY AND
THE MODEL INDEPENDENT OBSERVABLES FOR ABELIAN $Z^{\prime}$ SEARCH}
\author{A. V. Gulov
 \thanks{E-mail address: gulov@ff.dsu.dp.ua} and
 V. V. Skalozub
 \thanks{E-mail address: skalozub@ff.dsu.dp.ua}}
\address{Dniepropetrovsk State University, Dniepropetrovsk,
 320625 Ukraine}
\date{\today}
\maketitle
\begin{abstract}
The observables useful for the model independent search for
signals of the abelian $Z^\prime$ in the processes $e^{+}e^{-}\to
{\bar f}f$ are introduced. They are based on the renormalization
group relations between the $Z^\prime$ couplings to the Standard
Model particles developed recently and  extend the variables
suggested by Osland, Pankov and Paver. The bounds on the values of
the observables at the center-of-mass energy $\sqrt{s} = 500$GeV
are derived.
\end{abstract}

%
%

\section{Introduction}\label{sec:intr}

The abelian $Z^\prime$-boson with the mass much larger than the
$W$-boson mass ($m_{Z^\prime}\gg m_W$) is predicted by a number of
extensions of the Standard Model (SM) of elementary particles
\cite{leike}. At the current energies $\sim m_W$ the $Z^\prime$ is
decoupled. It can be described by the model based on the effective
gauge group $SU(2)_L\times U(1)_Y\times{\tilde U}(1)$ which is
assumed to be a low energy remnant of some unknown underlying
theory (GUT, for example). However, the $Z^\prime$ would be light
enough to give the first signal in future experiments.

Due to the lower mass limit from the Tevatron, $m_{Z^\prime}>{\cal
O}(500)$GeV, only the `indirect' $Z^\prime$ manifestations caused
by virtual heavy states can be searched for at the energies of the
present day accelerators. In general, one is not able to estimate
the magnitude of the $Z^\prime$ signal because of the unknown
couplings and the mass.
However, numerous strategies to evidence manifestations of the
$Z^\prime$ in experiments at high energy $e^+ e^-$ and hadronic
colliders have been developed. The analysis can be done with or
without assumptions on the specific underlying theory containing
the $Z^\prime$. Hence, there are model dependent and model
independent variables allowing to detect the $Z^\prime$.

One of the model independent approaches useful in searching for
the $Z^\prime$ signal in the leptonic processes $e^+ e^-\to \ell^+
\ell^-$ has been proposed in Ref. \cite{pankov}. The basic idea
was to replace the standard observables, the total cross section
$\sigma_T$ and the forward-backward asymmetry $A_{FB}$, by the new
set of variables defined as the differences of the cross sections
integrated over suitable ranges of polar angle $\theta$
\begin{equation}\label{int:1}
 \sigma_\pm \equiv
  \pm\int\limits_{\mp z^\ast}^1 \frac{d\sigma}{d\cos\theta}d\cos\theta
  \mp\int\limits_{-1}^{\mp
  z^\ast}\frac{d\sigma}{d\cos\theta}d\cos\theta.
\end{equation}
Due to the SM values of the leptonic charges and the kinematic
properties of the fermionic currents they have chosen the value
$z^\ast =2^{2/3}-1=0.5874$ to make the leading order deviations
from the SM predictions $\Delta\sigma_\pm$ dependent on the
combinations $v^e_{Z^\prime} v^\ell_{Z^\prime}\pm a^e_{Z^\prime}
a^\ell_{Z^\prime}$, where $v^f_{Z^\prime}$ and $a^f_{Z^\prime}$
parameterize the vector and the axial-vector coupling of the
$Z^\prime$ to the fermion $f$. Therefore, assuming the lepton
universality one obtains the sign definite observable
$\Delta\sigma_+$.

Usually, the parameters describing at low energies the $Z^\prime$
coupling to the SM fermions (like $v^f_{Z^\prime}$ and
$a^f_{Z^\prime}$) are assumed to be arbitrary numbers which must
be fixed in experiments. However, this is not the case if the
renormalizability of the underlying theory is taken into account.
In Refs. \cite{Yaf}, \cite{Zpr} it has  been shown that, if one
uses the principles of the renormalization group (RG) and the
decoupling theorem the correlations between the parameters
describing interactions of light particles with heavy virtual
states of  new physics beyond the SM, can be derived. Most
important that the relations obtained, being the consequence of
the renormalizability formulated in the framework of scattering in
the external field, are independent of the specific underlying
(GUT) model.

In Ref. \cite{Zpr} the method was applied to find signals of the
heavy abelian $Z^\prime$ in the four-fermion scattering processes.
It was found that the renormalizability of the theory is resulted
in the following constraints on the $Z^\prime$ couplings to the SM
fermions:
\begin{equation} \label{int:2}
v^{f^\prime}_{Z^\prime} - a^{f^\prime}_{Z^\prime}
 = v^f_{Z^\prime} - a^f_{Z^\prime},
 ~~ a^f_{Z^\prime} = I^f_3 Y_{\phi},
\end{equation}
where $f^\prime$ denotes the iso-partner of $f$ ($\ell^\prime =
\nu_\ell$, $\nu^\prime_\ell = \ell$, $q^\prime_d = q_u$,
$q^\prime_u = q_d$, where $\ell = e, \mu, \tau$ stands for
leptons), $I^f_3$ is the third component of the weak isospin and
$Y_\phi$ is the hypercharge parameterizing the $Z^\prime$ coupling
to the SM scalar doublet. Since the parameters of various
fermionic processes are appeared to be correlated, one could
expect that it is possible to introduce the specific observables
sensitive to the $Z^\prime$ manifestations. In the present paper
we propose such the observables which generalize the variables
$\sigma_\pm$ (\ref{int:1}).

The content is  as follows. In Sect. \ref{sec:nc} the
structure of neutral currents induced by the $Z^\prime$-boson as
well as the $Z$-$Z^\prime$ mixing is briefly discussed. Employing
the relations (\ref{int:2}) between the $Z^\prime$ parameters the
optimal observables for searching for signals of the heavy abelian
$Z^\prime$ are constructed in Sect. \ref{sec:observ}. In Sect.
\ref{sec:exp} the experimental bounds on the observables are
predicted. The obtained results are summarized in Sect.
\ref{sec:discuss}.

%
%

\section{Neutral currents and vector boson mixing}\label{sec:nc}

Considering the interactions of the $Z^\prime$ with the SM
particles, one has to conclude that at low energies, $E\ll
m_{Z^\prime}$, the renormalizable interactions are to be dominant.
The terms of the non-renormalizable type (for example, $\sim
(\partial_\mu Z^\prime_\nu -\partial_\nu Z^\prime_\mu)
{\bar\psi}\sigma_{\mu\nu} \psi$), being generated at GUT (or some
intermediate $\Lambda^{GUT}>\Lambda^\prime>m_{Z^\prime}$) mass
scale, are suppressed by the factors $1/\Lambda^{GUT}$,
$1/\Lambda^\prime$ and can be neglected. Thus, the interaction of
the $Z^\prime$ boson with the fermionic currents can be specified
by the effective Lagrangian
\begin{equation}\label{nc:1}
  {\cal L}_{NC} =eA_\mu J^\mu_A +g_Z Z_\mu J^\mu_Z
   +g_{Z^\prime}Z^\prime_\mu J^\mu_{Z^\prime},
\end{equation}
where $A, Z, Z^\prime$ are the photon, the $Z$- and
$Z^\prime$-bosons, respectively, $e=\sqrt{4\pi\alpha}$,
$g_Z=e/\sin \theta_W \cos \theta_W$, and $g_{Z^\prime}$ stands for
the ${\tilde U}(1)$ coupling constant. $\theta_W$ denotes the SM
value of the Weinberg angle ($\tan \theta_W =g^\prime/g$, where
the charges $g, g^\prime$ correspond to the gauge groups
$SU(2)_L$, $U(1)_Y$, respectively). The neutral currents can be
parameterized as
\begin{equation}\label{nc:2}
 J^\mu_V =\sum\limits_f {\bar f}\gamma^\mu
  \left( v^f_V +a^f_V \gamma^5 \right)f,
\end{equation}
with $V\equiv A,Z,Z^\prime$. The vector and the axial-vector
couplings of the vector boson $i$ to the fermion $f$ are
\begin{equation}\label{nc:3a}
  v^f_A =Q_f,\quad a^f_A=0,
\end{equation}
\begin{eqnarray}\label{nc:3b}
 v^f_Z&=&
  \left(\frac{I^f_3}{2} -Q_f \sin^2 \theta_W\right)\cos \theta_0
  +\frac{g_{Z^\prime}}{g_Z} Y^v_f \sin \theta_0,
 \quad\nonumber\\
 a^f_Z&=&
  -\frac{I^f_3}{2}\cos \theta_0
  +\frac{g_{Z^\prime}}{g_Z} Y^a_f \sin \theta_0,
\end{eqnarray}
\begin{eqnarray}\label{nc:3c}
 v^f_{Z^\prime}&=&
  Y^v_f \cos \theta_0
  -\frac{g_Z}{g_{Z^\prime}}
   \left(\frac{I^f_3}{2}-Q_f \sin^2 \theta_W\right) \sin \theta_0,
 \quad\nonumber\\
 a^f_{Z^\prime}&=&
  Y^a_f\cos{\theta}_{0}
  +\frac{g_Z}{g_{Z^\prime}}\frac{I^f_3}{2}\sin \theta_0,
\end{eqnarray}
where $Q_f$ is the fermion charge in the positron charge units.
The constants $Y^v_f$ and $Y^a_f$ parameterize the vector and the
axial-vector coupling of the fermion $f$ to the ${\tilde U}(1)$
symmetry eigenstate, whereas $\theta_0$ is the mixing angle
relating the mass eigen states $Z_\mu, Z^\prime_\mu$ to the
massive neutral components of the $SU(2)_L\times U(1)_Y$ and the
${\tilde U}(1)$ gauge fields, respectively. Its value can be
determined from the relation \cite{sirlin}
\begin{equation}\label{nc:4}
 \tan^2 \theta_0 =\frac{m^2_W/ \cos^2 \theta_W -m^2_Z}
  {m^2_{Z^\prime} -m^2_W/ \cos^2 \theta_W}.
\end{equation}

Because of the mixing between the $Z$ and $Z^\prime$ bosons the
mass $m_Z$ differs from the SM value $m_W/\cos \theta_W$ by the
small quantity of order $m^2_W/ m^2_{Z^\prime}$ \cite{Zpr}
\begin{equation}\label{nc:6}
  m^2_Z =\frac{m^2_W}{{\cos^2}{\theta_W}}
   \left(1-\frac{4{g^2_{Z^\prime}}{Y^2_\phi}}{g^2}
    \frac{m^2_W}{m^2_{Z^\prime} -m^2_W/{\cos^2}{\theta_W}}\right),
\end{equation}
and, as a consequence, the parameter $\rho\equiv m^2_W/ m^2_Z
\cos^2 \theta_W >1$. Therefore, the mixing angle $\theta_0$ is
also small $\theta_0 \simeq\tan \theta_0 \simeq\sin \theta_0 \sim
m^2_W/ m^2_{Z^\prime}$.

The difference $m^2_Z -m^2_W/ \cos^2 \theta_W$ is negative and
completely determined by the $Z^\prime$ coupling to the scalar
doublet. Thus, constraints on the $Z^\prime$ interaction with the
scalar field can be obtained by experimental detecting this
observable:
\begin{equation}\label{nc:7}
  \frac{g^2_{Z^\prime} Y^2_\phi}{m^2_{Z^\prime}}=
  \left(1- \frac{m^2_Z \cos^2 \theta_W}{m^2_W}\right)\frac{g^2}{4m^2_W}
  +O\left(\frac{m^4_W}{m^4_{Z^\prime}}\right) .
\end{equation}

As it has been proven in Ref.\cite{Zpr}, the following relations
hold for the constants $Y^v_f$ and $Y^a_f$
\begin{equation}\label{nc:5}
Y^L_{f^\prime}=Y^L_f,\quad Y^a_f=I^f_3 Y_\phi,
\end{equation}
where $Y^L_f\equiv Y^v_f -Y^a_f$, $Y^R_f\equiv Y^v_f +Y^a_f$, and
$Y_\phi$ is the hypercharge parameterizing the coupling of the SM
scalar doublet to the vector boson associated with the ${\tilde
U}(1)$ symmetry. As it is noted in Sect.
\ref{sec:intr}, the notation $f^\prime$ stands for the iso-partner
of $f$.

In fact, the relation (\ref{nc:5}) means that the $Z^\prime$
couplings to the SM axial currents have the universal absolute
value, if a single light scalar doublet exists. Among the four
values $Y^v_f$, $Y^a_f$, $Y^v_{f^\prime}$, $Y^a_{f^\prime}$
parameterizing interaction of the $Z^\prime$-boson with the
$SU(2)$ fermionic isodoublet only one is independent. The rest
ones can be expressed through it and the hypercharge $Y_\phi$ of
the $Z^\prime$ coupling to the SM scalar doublet. If the
hypercharges are treated as unknown parameters these relations are
to be taken into account in order to preserve the gauge symmetry
\cite{Zpr}. The relations (\ref{nc:5})  also show that the
fermion and the scalar sectors of the new physics are strongly
connected. As a result, the couplings of the $Z^\prime$-boson to
the SM axial currents are completely determined by its interaction
with the scalar fields. Therefore, one is able to predict the $Z^\prime$
coupling to the SM axial currents by measuring the $\rho$
parameter. When the $Z^\prime$ does not interact with the scalar
doublet, the $Z$-boson mass is to be identical to its SM value.
In this case the $Z^\prime$ couplings to the axial currents are produced
 by loops and to be suppressed by the additional small factor $g^2/16 \pi^2$.


%
%

\section{The observables}\label{sec:observ}

\begin{figure}
\begin{center}
 \epsfxsize=0.3\textwidth
 \epsfbox[0 0 600 600]{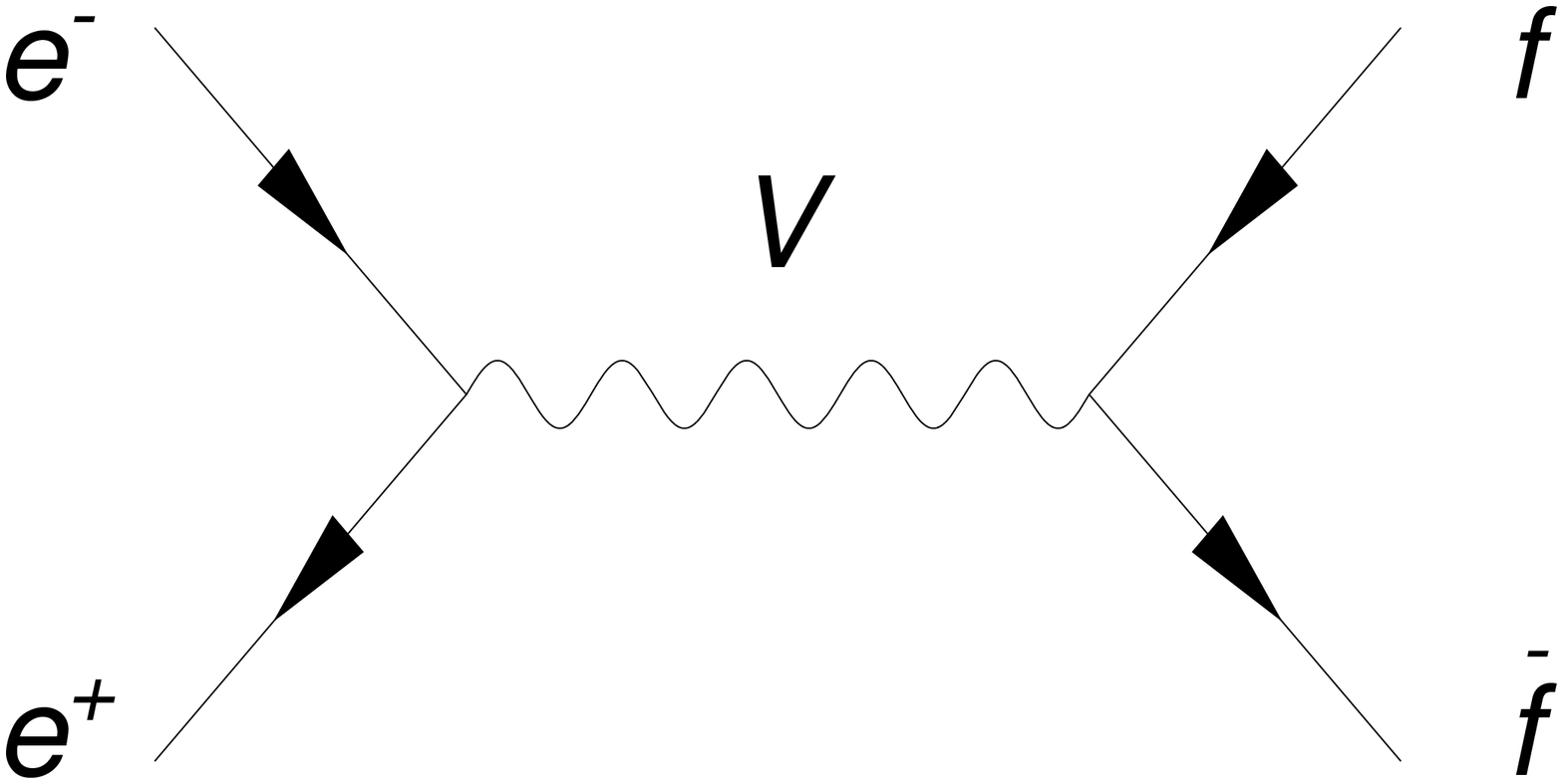}
 \caption{The amplitude $T_{V}$ of the process
 $e^{+}e^{-}\to V^{\ast}\to{\bar f}f$ at the Born level.}
 \label{fig:1}
 \end{center}
\end{figure}

In the present section we consider the electron-positron annihilation
into fermion pairs $e^+ e^- \to V^\ast \to{\bar f}f$ for energies
 $\sqrt{s}\sim 500$GeV. In  this case all the fermions except
for the $t$-quark can be treated as massless particles $m_f\sim
0$. The $Z^\prime$-boson existence causes the deviations ($\sim
m^{-2}_{Z^\prime}$) of the cross section from its SM value:
\begin{equation}\label{obs:1}
 \Delta\frac{d\sigma}{d\Omega}=
   \frac{d\sigma}{d\Omega}-\frac{d \sigma_{SM}}{d\Omega}
   =\frac{\mbox{Re}\left[T^\ast_{SM}\Delta T\right]}
   {32\pi s}+O(\frac{s^2}{m^4_{Z^\prime}}),
\end{equation}
with
\begin{equation}\label{obs:2}
 T_{SM}=T_A+\left.T_Z\right|_{\theta_0 =0},~~
 \Delta T =T_{Z^\prime}
  +\left.\frac{dT_Z}{d \theta_0}\right|_{\theta_0 =0}{\theta}_0,
\end{equation}
where $T_V$ denotes the Born amplitude of the process $e^+ e^- \to
V^\ast \to{\bar f}f$ with the virtual $V$-boson ($V=A,Z,Z^\prime$)
state in the $s$-channel (the corresponding diagram is shown in
Fig. \ref{fig:1}).

The quantity $\Delta d\sigma/d\Omega$ can be calculated in the form
\begin{eqnarray}\label{obs:3}
  \Delta\frac{d\sigma}{d\Omega}
  &=&\frac{\alpha I^f_3 N_f}{4\pi}\sum\limits_{\lambda,\xi}
   \frac{g^2_{Z^\prime} \zeta^{ef}_{\lambda\xi}}{m^2_{Z^\prime}}
    \left( |Q_f| +\chi (s)\left( \mbox{sgn} \lambda
    -\varepsilon\right)\right.
   \nonumber\\&&\times\left.
    \left( \mbox{sgn} \xi + |Q_f|(1-\varepsilon )-1\right)\right)
   {\left( z + \mbox{sgn} \lambda\xi \right)}^2,
\end{eqnarray}
where $N_f=3$ for quarks and $N_f=1$ for leptons,
$\lambda,\xi=L,R$ denotes the fermion helicity states, $\chi (s)={(16
\sin^2 \theta_W \cos^2 \theta_W (1-m^2_Z/s))}^{-1}$,
$z\equiv\cos\theta$ (where $\theta$ is the angle between the
incoming electron and the outgoing fermion), $\varepsilon\equiv
1-4 \sin^2 \theta_W \sim 0.08$, and
\begin{eqnarray}\label{obs:4}
  \zeta^{ef}_{\lambda\xi}&\equiv& Y^\lambda_e Y^\xi_f
   -\frac{m^2_W/{\cos}^2{\theta}_W}{s -m^2_Z}
     \left(
     Y_\phi Y^\xi_f
     \left( 2{\sin}^2{\theta}_W -\delta_{\lambda,L} \right)
     \right.\nonumber\\
     &&+\left.
     2 I^f_3 Y_\phi Y^\lambda_e
     \left( -2|Q_f|{\sin}^2{\theta}_W + \delta_{\xi,L} \right)
     \right),
\end{eqnarray}
with $\delta_{\lambda,\xi}=1$ when $\lambda=\xi$ and
$\delta_{\lambda,\xi}=0$ otherwise.

To discuss the physical effects caused by the relation
(\ref{nc:5}), let us introduce the observable
$\Delta\sigma\left(Z\right)$ defined as the difference of cross
sections integrated in suitable ranges of $\cos\theta$
\begin{equation}\label{obs:5}
  \sigma\left(Z\right)
  \equiv\int\limits_Z^1\frac{d\sigma}{dz}dz
   -\int\limits_{-1}^Z \frac{d\sigma}{dz}dz
\end{equation}
The value of $Z$ will be chosen later. Actually, this observable is
the generalized $\sigma_+$ of Ref. \cite{pankov} ($\sigma_+
=\sigma(-z^\ast)$). The two conventionally used observables, the
total cross section $\sigma_T$ and the forward-backward asymmetry
$A_{FB}$, can be obtained by special choice of $Z$
($\sigma_T=\sigma (-1)$, $A_{FB}=\sigma (0)/ \sigma_T$). In fact,
one can express $\sigma(Z)$ in terms of $\sigma_T$ and $A_{FB}$
\begin{equation}\label{obs:5a}
  \sigma\left(Z\right) =\sigma_T\left(
  A_{FB}\left(1-Z^2\right) -\frac{1}{4}Z\left(3+Z^2\right)\right).
\end{equation}

\begin{figure}
\begin{center}
 \epsfxsize=0.35\textwidth
 \epsfbox[0 0 600 600]{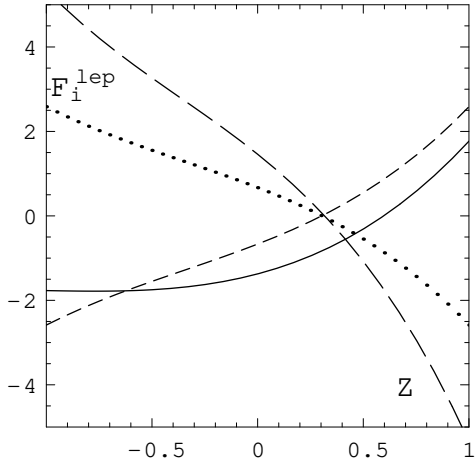}
 \caption{The leptonic functions
  $F^{\ell}_{0}$ (the solid curve),
  $F^{\ell}_{1}$ (the long-dashed curve),
  $F^{\ell}_{2}$ (the dashed curve) and
  $F^{\ell}_{3}$ (the dotted curve) at $\sqrt{s}=500$GeV.}
 \label{fig:2}
 \end{center}
\end{figure}

\begin{figure}
\begin{center}
 \epsfxsize=0.35\textwidth
 \epsfbox[0 0 600 600]{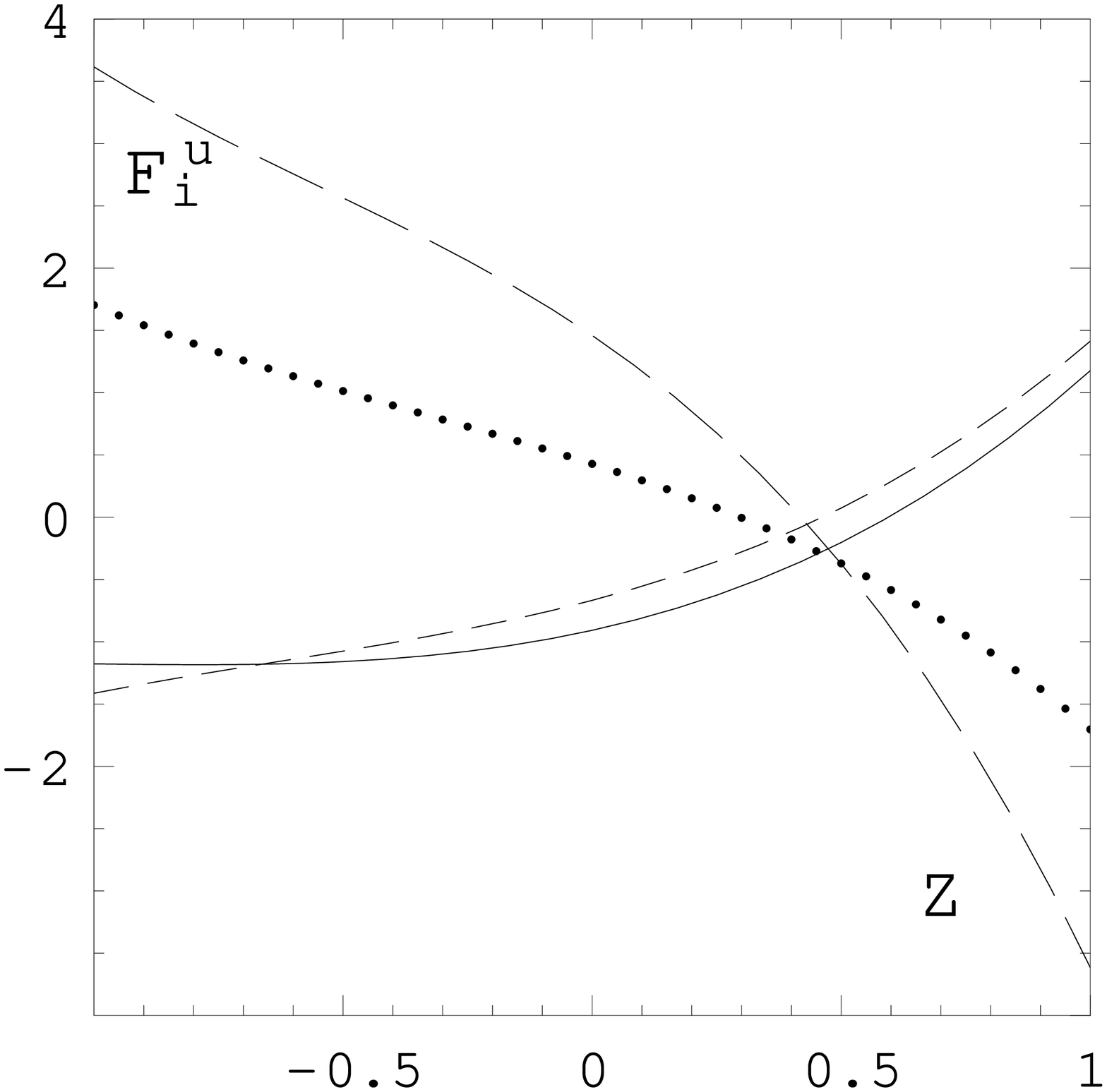}
 \caption{The quark functions ($q_{u}=u,c$)
  $F^{q_{u}}_{0}$ (the solid curve),
  $F^{q_{u}}_{1}$ (the long-dashed curve),
  $F^{q_{u}}_{2}$ (the dashed curve) and
  $F^{q_{u}}_{3}$ (the dotted curve) at $\sqrt{s}=500$GeV.}
 \label{fig:3}
 \end{center}
\end{figure}

\begin{figure}
\begin{center}
 \epsfxsize=0.35\textwidth
 \epsfbox[0 0 600 600]{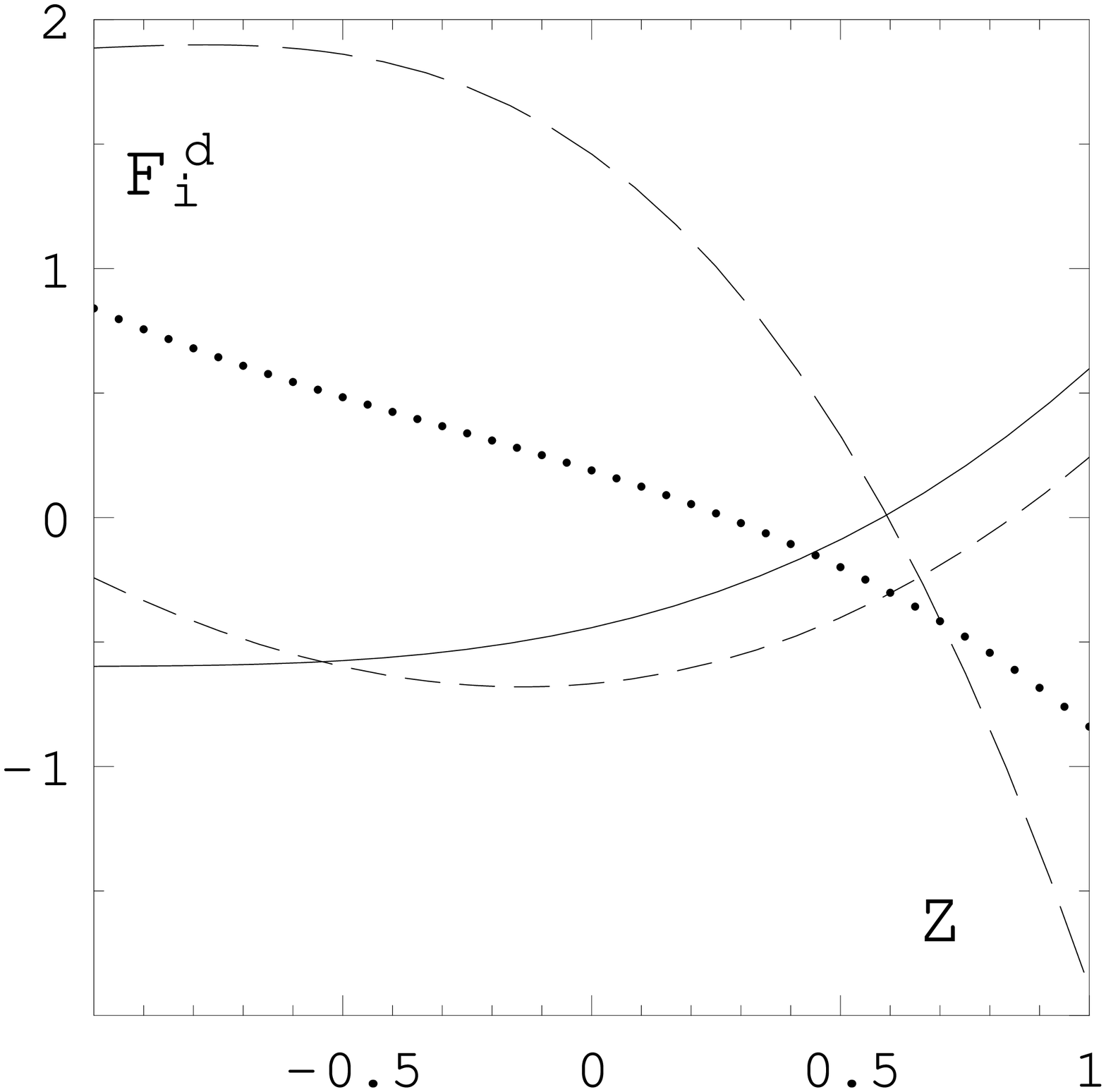}
 \caption{The quark functions ($q_{d}=d,s,b$)
  $F^{q_{d}}_{0}$ (the solid curve),
  $F^{q_{d}}_{1}$ (the long-dashed curve),
  $F^{q_{d}}_{2}$ (the dashed curve) and
  $F^{q_{d}}_{3}$ (the dotted curve) at $\sqrt{s}=500$GeV.}
 \label{fig:4}
 \end{center}
\end{figure}

Owing to the relations (\ref{nc:5}) the quantity
$\Delta\sigma\left(Z\right)\equiv
\sigma\left(Z\right)-{\sigma}_{SM}\left(Z\right)$ can be written
in the form
\begin{eqnarray}\label{obs:6}
 \Delta\sigma\left(Z\right)
 &=&\frac{\alpha N_f}{8}\frac{g^2_{Z^\prime}}{m^2_{Z^\prime}}
  \left(
  F^f_0(Z,s) Y^2_\phi +2 F^f_1(Z,s) I^f_3 Y^L_f Y^L_e\right.
 \nonumber\\
  &&+\left. 2 F^f_2(Z,s) I^f_3 Y^L_f Y_\phi
    + F^f_3(Z,s) Y^L_e Y_\phi \right).
\end{eqnarray}
The functions $F^f_i(Z,s)$ depend on the fermion type through the
$|Q_f|$, only. In Figs. \ref{fig:2}-\ref{fig:4} they are shown
as the functions of $Z$ for $\sqrt{s}=500$GeV. The leading
contributions to $F^f_i(Z,s)$
\begin{eqnarray}\label{obs:7}
  F^f_0(Z,s)&=&-\frac{4}{3}\left|Q_f\right|
   \left(1 -Z -Z^2 -\frac{Z^3}{3}\right)
  \nonumber\\&&+O\left(\varepsilon,\frac{m^2_Z}{s}\right),
  \nonumber\\
  F^f_1(Z,s)&=&\frac{4}{3}
   \left(1 -Z^2 -\left|Q_f\right|\left(3Z+Z^3\right)\right)
   \nonumber\\&&+O\left(\varepsilon,\frac{m^2_Z}{s}\right),
   \nonumber\\
  F^f_2(Z,s)&=&-\frac{2}{3}\left(1-Z^2\right)
   +\frac{2}{9}\left(3Z+Z^3\right)
   \nonumber\\
   &&\times\left(4\left|Q_f\right|-1\right)
      +O\left(\varepsilon,\frac{m^2_Z}{s}\right),
  \nonumber\\
  F^f_3(Z,s)&=&\frac{2}{3}\left|Q_f\right|
   \left(1-3Z-Z^2-Z^3\right)
  \nonumber\\&&+O\left(\varepsilon,\frac{m^2_Z}{s}\right)
\end{eqnarray}
are given by the $Z^\prime$ exchange diagram (the first term of
Eq.(\ref{obs:4})), since the contribution of the $Z$ exchange
diagram to $\Delta T$ (the second term of Eq.(\ref{obs:4})) is
suppressed by the factor $m^2_Z/s$.

From the Eqs. (\ref{obs:7}) one can see that the leading
contributions to the leptonic factors $F^\ell_1$, $F^\ell_2$,
$F^\ell_3$ are found to be proportional to the same polynomial in
$Z$. This is the characteristic feature of the leptonic functions
$F^\ell_i$ originated due to the kinematic properties of fermionic
currents and the specific values of the SM leptonic charges.
Therefore, it is possible to choose the value of $Z=Z^\ast$ which
switches off three leptonic factors $F^\ell_1$, $F^\ell_2$,
$F^\ell_3$ simultaneously. Moreover, the quark function $F^q_3$ in
lower order is proportional to the leptonic one  and therefore is
switched off, too. As is seen from Figs. \ref{fig:2}-\ref{fig:4},
the appropriate value of $Z^\ast$ is about $\sim 0.3$. By choosing
this value of $Z^\ast$ one can simplify Eq.(\ref{obs:6}). It is
also follows from Eq. (\ref{obs:6}) that neglecting the factors
$F^\ell_1$, $F^\ell_2$, $F^\ell_3$ one obtains the sign definite
quantity $\Delta \sigma_\ell (Z^\ast)$.

Comparing the observable $\Delta \sigma_\ell(Z^\ast)$ with $\Delta
\sigma_+ =\Delta \sigma_\ell(-0.5874)$ or $\Delta \sigma_-
=-\Delta \sigma_\ell(0.5874)$ one can to conclude that the
sign of the variables $\Delta \sigma_\pm$ is completely
undetermined in the case of arbitrary leptonic couplings
$Y^L_\ell$. Therefore, in order to predict the sign of the
observables $\Delta \sigma_\pm$ one has to assume the additional
restriction such as the lepton universality.

\begin{figure}
\begin{center}
 \epsfxsize=0.35\textwidth
 \epsfbox[0 0 600 600]{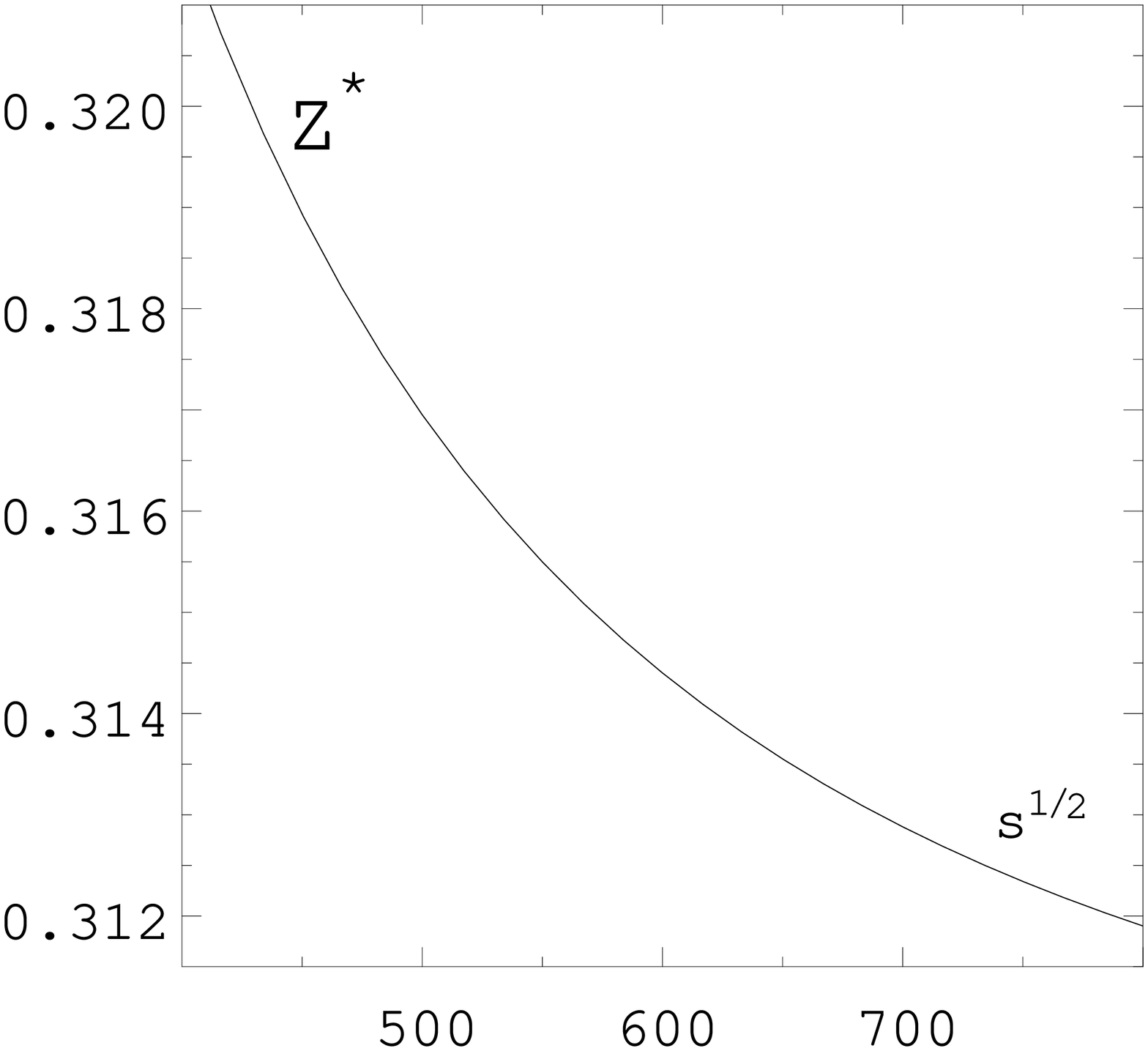}
 \caption{$Z^\ast$ as the function of $\sqrt{s}$(GeV).}
 \label{fig:5}
 \end{center}
\end{figure}

Let the value of $Z^\ast$ in Eq.(\ref{obs:5}) is determined from
the relation
\begin{equation}\label{obs:8}
  F^\ell_1\left(Z^\ast,s\right)=0.
\end{equation}
The solution $Z^\ast(s)$ is shown in Fig \ref{fig:5}. As is seen,
$Z^\ast$ decreases from 0.3170 at $\sqrt{s}=500$GeV to 0.3129 at
$\sqrt{s}=700$GeV. Table I demonstrates the corresponding behavior
of the functions $F^f_i(Z^\ast,s)$. Since $F^f_i(Z^\ast,s)$ depend
on the center-of-mass energy through the small quantity $m^2_Z/s$,
the order of the shifts is about 3\%. Therefore, in what follows
the value of $\sqrt{s}$ is taken to be 500GeV.

Assuming $Y^L_e\sim Y^L_\ell\sim Y_\phi\sim Y^L_u\sim 1$, one can
derive
\begin{eqnarray}\label{obs:9}
 \Delta \sigma_\ell\left(Z^\ast\right)
 &=&-0.10\frac{\alpha g^2_{Z^\prime} Y^2_\phi}{m^2_{Z^\prime}}
  \left(1 +O\left(0.04\right)\right),
  \nonumber\\
 \Delta \sigma_{q_u}\left(Z^\ast\right)
 &=&1.98\Delta \sigma_\ell\left(Z^\ast\right)
  +0.32\frac{\alpha g^2_{Z^\prime} Y_\phi}{2m^2_{Z^\prime}}
 \nonumber\\&&
  \times\left( \left( Y^L_e/ Y_\phi-0.6\right)
   Y^L_{q_u}+O\left(0.07\right)\right),
 \nonumber\\
 \Delta \sigma_{q_d}\left(Z^\ast\right)
 &=&0.94 \Delta \sigma_\ell\left(Z^\ast\right)
  -0.32\frac{\alpha g^2_{Z^\prime} Y_\phi}{m^2_{Z^\prime}}
 \nonumber\\&&
  \times\left(
   \left( Y^L_e/ Y_\phi-0.6\right) Y^L_{q_u}
   +O\left(0.08\right) \right),
\end{eqnarray}

Hence it is seen, that the observable $\Delta \sigma_\ell(Z^\ast)$
is negative. Moreover, it can be written in terms of the $\rho$
parameter using Eq.(\ref{nc:7}):
\begin{equation}\label{obs:10}
 \Delta \sigma_\ell \left(Z^\ast\right)
 \simeq
  0.10\frac{\alpha g^2 \left(1-\rho\right)}{4m^2_W\rho}<0.
\end{equation}


One also can construct the sign definite observable for
quarks of the same generation. As it follows from Eq.
(\ref{obs:9}),
%
\begin{equation}\label{obs:13}
 \Delta \sigma_{q_u}\left(Z^\ast\right)
 +0.5\Delta \sigma_{q_d}\left(Z^\ast\right)
 \simeq 2.45\Delta \sigma_\ell\left(Z^\ast\right)<0.
\end{equation}
Hence it follows, that the values of $\Delta \sigma_{q_u}(Z^\ast)$
and $\Delta \sigma_{q_d}(Z^\ast)$ in the $\Delta
\sigma_{q_u}(Z^\ast)$ -- $\Delta \sigma_{q_d}(Z^\ast)$ plane have
to be at the line crossing axes at the points $\Delta
\sigma_{q_u}(Z^\ast)=2.45\Delta \sigma_\ell(Z^\ast)$ and $\Delta
\sigma_{q_d}(Z^\ast)=4.9\Delta \sigma_\ell(Z^\ast)$, respectively.
It also follows from Eq.(\ref{obs:10}) that the observable $\Delta
\sigma_{q_u}(Z^\ast) +0.5\Delta \sigma_{q_d}(Z^\ast)$ is negative.

Thus, the dependencies (\ref{int:2}) between the $Z^\prime$
couplings to SM fermions allows to construct three negative valued
observables, $\Delta \sigma_\ell (Z^\ast)$, $1-\rho$ and $\Delta
\sigma_{q_u}(Z^\ast)+0.5\Delta \sigma_{q_d}(Z^\ast)$, which are
correlated by Eqs. (\ref{obs:10})-(\ref{obs:13}). These
observables are the most general model independent ones which can
be introduced without any assumptions such as the lepton or quark
universality.

\section{The experimental constraints on the observables}\label{sec:exp}

The present day  experimental data constrain the magnitude of the
four-fermion contact interactions allowing to derive bounds on the
$Z^\prime$ coupling to the axial currents and, consequently, on
the observables introduced in the previous section. Our analysis
is based on the data presented in Refs. \cite{bounds} where the
study of the experimental bounds on the lepton-quark four-fermion
contact couplings has been performed. In general, the
contributions of new physics beyond the SM to the considered
therein processes (the atomic parity violation experiment as well
as the electron-nucleus, muon-nucleus and $\nu_\mu$-nucleon
scattering experiments) are described by 20 parameters, namely,
\begin{equation}\label{exp:1}
  \eta^{\ell q}_{\lambda\xi}
  \equiv -\frac{g^2_{Z^\prime}Y^\lambda_\ell
  Y^\xi_q}{m^2_{Z^\prime}},\quad
  \eta^{\nu_\mu q}_{L\xi}
  \equiv -\frac{g^2_{Z^\prime}Y^L_{\nu_\mu}
  Y^\xi_q}{m^2_{Z^\prime}},
\end{equation}
where $\ell=e,\mu$; $q=u,d$ and $\lambda,\xi=L,R$. In order to
reduce the number of independent $\eta^{\ell q}_{\lambda\xi}$ one
usually assumes the $SU(2)_L$ invariance and the lepton
universality. As a result, six variables (for example, $\eta^{\ell
u}_{LL}$, $\eta^{\ell u}_{LR}$, $\eta^{\ell d}_{LR}$, $\eta^{\ell
u}_{RL}$, $\eta^{\ell u}_{RR}$, $\eta^{\ell d}_{RR}$) can be
chosen as the basis.

However, the number of independent $\eta^{\ell q}_{\lambda\xi}$
can be further decreased by employing the correlations
(\ref{int:2}). In this case it is useful to introduce the
couplings $\eta^{\ell q}_{AA}$, $\eta^{\ell q}_{LA}$, $\eta^{\ell
q}_{AL}$ parameterizing the four-fermion interactions between the
left-handed and the axial-vector currents. These couplings are the
linear combinations of the variables (\ref{exp:1}):
\begin{eqnarray}\label{exp:1a}
  \eta^{\ell q}_{AA}&&
   \equiv\eta^{\ell q}_{RR} -\eta^{\ell q}_{RL}
   -\eta^{\ell q}_{LR} +\eta^{\ell q}_{LL},
  \nonumber\\
  \eta^{\ell q}_{LA}&&
   \equiv \eta^{\ell q}_{LR} -\eta^{\ell q}_{LL},
  \nonumber\\
  \eta^{\ell q}_{AL}&&
   \equiv\eta^{\ell q}_{RL} -\eta^{\ell q}_{LL}.
\end{eqnarray}
As it follows from Eq. (\ref{int:2}), one has six independent
parameters
\begin{eqnarray}\label{exp:2}
  \eta^{eu}_{AA}&&= \frac{g^2_{Z^\prime}Y^2_\phi}{4m^2_{Z^\prime}},
  \quad
  \eta^{\ell u}_{LA}= -\frac{g^2_{Z^\prime}
   Y^L_\ell Y_\phi}{2m^2_{Z^\prime}},
  \nonumber\\
  \eta^{eu}_{AL}&&= \frac{g^2_{Z^\prime}
   Y^L_u Y_\phi}{2m^2_{Z^\prime}},
  \quad
  \eta^{\ell u}_{LL}= -\frac{g^2_{Z^\prime}
   Y^L_\ell Y^L_u}{m^2_{Z^\prime}},
\end{eqnarray}
which can be used as the basis.

The experiment constrains the specific linear combinations of the
variables $\eta^{\ell q}_{\lambda\xi}$ (see Ref. \cite{bounds}).
Introducing the normalized couplings:
\begin{eqnarray}\label{exp:2a}
 &&\Delta C^\ell_{1q}=-\frac{1}{2\sqrt{2}G_F}
  \left(\eta^{\ell q}_{RR}-\eta^{\ell q}_{LR}
   +\eta^{\ell q}_{RL}-\eta^{\ell q}_{LL}\right),
 \nonumber\\&&
 \Delta C^\ell_{2q}=-\frac{1}{2\sqrt{2}G_F}
  \left(\eta^{\ell q}_{RR}+\eta^{\ell q}_{LR}
   -\eta^{\ell q}_{RL}-\eta^{\ell q}_{LL}\right),
 \nonumber\\&&
 \Delta C^\ell_{3q}=-\frac{1}{2\sqrt{2}G_F}
  \left(\eta^{\ell q}_{RR}-\eta^{\ell q}_{LR}
   -\eta^{\ell q}_{RL}+\eta^{\ell q}_{LL}\right),
 \nonumber\\&&
 \Delta q_L=-\frac{1}{2\sqrt{2}G_F}\eta^{\nu_\mu q}_{LL},
 \nonumber\\&&
 \Delta q_R=-\frac{1}{2\sqrt{2}G_F}\eta^{\nu_\mu q}_{LR},
\end{eqnarray}
where $G_F$ is the Fermi constant, one can write down the
experimental bounds as follows:
\begin{eqnarray}
 &&2\Delta C^e_{1u} -\Delta C^e_{1d} = 0.217 \pm 0.26;
 \nonumber\\&&
 2\Delta C^e_{2u} -\Delta C^e_{2d} =-0.765 \pm 1.23;
 \label{exp:2b}\\&&
 2\Delta C^\mu_{3u}-\Delta C^\mu_{3d}=-1.51\pm 4.9;
 \nonumber\\&&
 2\Delta C^\mu_{2u}-\Delta C^\mu_{2d}=1.74\pm 6.3;
 \label{exp:2c}\\&&
 \Delta C^e_{1u}+\Delta C^e_{1d}=0.0152\pm 0.033;
 \label{exp:2d}\\&&
 -2.73\Delta C^e_{1u}+0.65\Delta C^e_{1d}
  -2.19\Delta C^e_{2u}+2.03\Delta C^e_{2d}
 \nonumber\\&&\quad =-0.065\pm 0.19;
 \label{exp:2e}\\&&
 376\Delta C^e_{1u}+422\Delta C^e_{1d}=0.96\pm 0.92;
 \nonumber\\&&
 572\Delta C^e_{1u}+658\Delta C^e_{1d}=1.58\pm 4.2;
 \label{exp:2f}\\&&
 \Delta u_L=-0.0032\pm 0.0169;
 \nonumber\\&&
 \Delta u_R=-0.0084\pm 0.0251;
 \nonumber\\&&
 \Delta d_L=0.002\pm 0.0136;
 \nonumber\\&&
 \Delta d_R=-0.0109\pm 0.0631
 \label{exp:2g}.
\end{eqnarray}
Eqs. (\ref{exp:2b})-(\ref{exp:2g}) represent the results of SLAC
$e-D$, CERN $\mu -C$, Bates $e-C$, Mainz $e-Be$, the atomic parity
violation and the $\nu_\mu$-nucleon scattering experiments,
respectively \cite{bounds}. They determine the allowed region in
the space of $\eta$- parameters.

By employing Eqs. (\ref{exp:1a}), (\ref{exp:2b})-(\ref{exp:2g}) it
is easy to obtain the bounds on the quantities (\ref{exp:2}):
\begin{eqnarray}\label{exp:3}
 0 &<&\eta^{eu}_{AA}< 0.114\mbox{TeV}^{-2},
  \nonumber\\
 -0.018\mbox{TeV}^{-2}&<&\eta^{eu}_{AL}<0.006\mbox{TeV}^{-2},
  \nonumber\\
 -0.437\mbox{TeV}^{-2}&<&\eta^{\mu u}_{LA}<0.661\mbox{TeV}^{-2},
  \nonumber\\
 -0.667\mbox{TeV}^{-2}&<&\eta^{eu}_{LA}<0.238\mbox{TeV}^{-2},
  \nonumber\\
 -0.423\mbox{TeV}^{-2}&<&\eta^{\mu u}_{LL}<0.358\mbox{TeV}^{-2}
\end{eqnarray}
The first of the relations (\ref{exp:3}) gives possibility to
determine the allowed magnitude of the observables
$\Delta{\sigma}_{\ell}(Z^\ast)$ and $\Delta \sigma_{q_u}(Z^\ast)
 +0.5\Delta \sigma_{q_d}(Z^\ast)$
\begin{eqnarray}\label{exp:4}
 -0.13\mbox{pb}&<&\Delta{\sigma}_\ell\left(Z^\ast\right)<0,
 \nonumber\\
 -0.32\mbox{pb}&<&\Delta \sigma_{q_u}\left(Z^\ast\right)
 +0.5\Delta \sigma_{q_d}\left(Z^\ast\right)<0.
\end{eqnarray}
Thus, the signals of the abelian $Z^\prime$ must respect the above
relations if the low energy physics is described by the minimal
SM.


It is worth to compare the first of Eqs. (\ref{exp:4}) based on
the analysis of the lepton-quark interactions with the direct
constraints derived from the experiments on the $e^+ e^- \to
\ell^+ \ell^-$ scattering. Introducing the normalized $Z^\prime$
coupling to the axial leptonic current:
\begin{equation}\label{exp:5}
  \left| A_\ell\right|=
  \sqrt{\frac{g^2_{Z^\prime}Y^{a2}_\ell m^2_Z}{4\pi m^2_{Z^\prime}}},
\end{equation}
one can write down the bounds obtained from the processes $e^+ e^-
\to \ell^+ \ell^-$ as follows $|A_\ell|<0.025$ (see Fig. 2.7 of
Ref. \cite{leike}). However, the relations (\ref{exp:4}) lead to
the constraint $|A_\ell|<0.0087$. Thus, the bounds (\ref{exp:4})
are about nine times stronger than ones derived from the analysis
of the pure leptonic interactions.

%
%

\section{Discussion}\label{sec:discuss}

In the lack of reliable information on the model describing
physics beyond the SM  and predicting $Z^\prime$ boson it is of
great importance to find the model independent variables to search
for this particle. In this regard, it could be useful to employ
the method of Ref. \cite{Zpr} which gives possibility to reduce
the number of unknown numbers parameterizing effects on new heavy
virtual particles.

In Ref. \cite{pankov} the observables ${\sigma}_{\pm}$ alternative
to the familiar $\sigma_T$ and $A_{FB}$ and perspective in
searching for the abelian $Z^{\prime}$ signal in the leptonic
processes $e^{+}e^{-}\to {\ell}^{+}{\ell}^{-}$ were proposed. It
was pointed out that the sign of the ${\sigma}_{+}$ deviation from
the SM value can be uniquely predicted when the lepton
universality is assumed. The sign remains to be unspecified in the
case of an arbitrary interaction of $Z^\prime$ with leptons.

The observables introduced in the present paper are the extension
of that in Ref. \cite{pankov} with the specified above choice of
the boundary angle. The observable $\Delta{\sigma}_{\ell}(Z^\ast)$
is the negative defined quantity even in the case when the lepton
universality is not assumed. Moreover, the observable
$\Delta{\sigma}_{q_{u}}(Z^\ast)
+0.5\Delta{\sigma}_{q_{d}}(Z^\ast)$ constructed for the quarks of
the same generation is proportional to the leptonic one
$\Delta{\sigma}_{\ell}(Z^\ast)$ being negative, too.

As it was mentioned in section \ref{sec:observ}, the magnitude of
the leptonic observable $\Delta{\sigma}_{\ell}(Z^\ast)$ is
determined by means of the $\rho$ parameter. Thus, the
$Z^{\prime}$ contributions to the quantities
$1-\rho=1-m^{2}_{W}/m^{2}_{Z}{\cos}^{2}{\theta}_{W}$,
$\Delta{\sigma}_{\ell}(Z^\ast)$ and
$\Delta{\sigma}_{q_{u}}(Z^\ast)
+0.5\Delta{\sigma}_{q_{d}}(Z^\ast)$ have to be negative allowing
to detect the $Z^{\prime}$ signal.

As the corollary of our analysis we note that the introduced
observables allow one to identify (or to discard) the abelian
$Z^{\prime}$ effects when the low energy physics is described by
the minimal SM. Therefore, they can usefully complement the
conventional analysis of $Z^{\prime}$ couplings based on the
observables $\sigma_T$ and $A_{FB}$.

The authors thank G.C. Cho for information on the results in Refs.
\cite{bounds}.

%
%

\newpage
\begin{table}
\begin{center}
\label{table:1} \caption{Energy dependence of $F^f_i(Z^\ast)$.}
\begin{tabular}{l l l l}
 $\sqrt{s}$,GeV & 500 & 600 & 700 \\
 \hline\\
 $Z^\ast$ & 0.3170 & 0.3144 & 0.3129 \\
 \hline\\
 $F^\ell_0 \left(Z^\ast\right)$ & -0.8012 & -0.7889 & -0.7815 \\
 $F^\ell_2 \left(Z^\ast\right)$ & 0.0346 & 0.0341 & 0.0338 \\
 $F^\ell_3 \left(Z^\ast\right)$ & -0.0346 & -0.0341 & -0.0338 \\
 $F^{q_u}_0\left(Z^\ast\right)$ & -0.5277 & -0.5216 & -0.5179 \\
 $F^{q_u}_1\left(Z^\ast\right)$ & 0.4250 & 0.4215 & 0.4194 \\
 $F^{q_u}_2\left(Z^\ast\right)$ & -0.2532 & -0.2499 & -0.2479 \\
 $F^{q_u}_3\left(Z^\ast\right)$ & -0.0331 & -0.0296 & -0.0276 \\
 $F^{q_d}_0\left(Z^\ast\right)$ & -0.2513 & -0.2522 & -0.2527 \\
 $F^{q_d}_1\left(Z^\ast\right)$ & 0.8500 & 0.8430 & 0.8388 \\
 $F^{q_d}_2\left(Z^\ast\right)$ & -0.5410 & -0.5339 & -0.5297 \\
 $F^{q_d}_3\left(Z^\ast\right)$ & -0.0362 & -0.0282 & -0.0235
\end{tabular}
\end{center}
\end{table}

%
%
\newpage
\newpage

\end{document}